\documentclass[twocolumn]{aastex631}

\usepackage{longtable} 
\usepackage{graphicx}
\usepackage{subfigure}
\usepackage{multirow}

\def \mgii  {Mg\,{\sc ii}}
\def \cii     {C\,{\sc ii}}

\def \siiv   	{Si\,{\sc iv}}

\def \six   	{Si\,{\sc x}}

\def \caii   	{Ca\,{\sc ii}}

\def \feix		{Fe\,{\sc ix}}

\def \fexviii  	{Fe\,{\sc xviii}}

\def \fexxi  	{Fe\,{\sc xxi}}

\def \iris   	{{\it IRIS}}

\def \sdo   	{{\it SDO}}
\def \hinode   	{{\it Hinode}}
\def \sst   	{{\it SST}}
\def \arcsec 	{\hbox{$^{\prime\prime}$}}

\def\kms{\hbox{km$\;$s$^{-1}$}}
\def\Halpha{\mbox{H\hspace{0.1ex}$\alpha$}}

\received{}

\submitjournal{ApJ}

\shorttitle{High resolution chromospheric observations of nanoflares}
\shortauthors{Testa et al.}

\begin{document}

\title{High Resolution Observations of the Low Atmospheric Response to Small Coronal Heating Events in an Active Region Core}
 
\correspondingauthor{Paola Testa}
\email{ptesta@cfa.harvard.edu}

\author[0000-0002-0405-0668]{Paola Testa}
\affil{Harvard-Smithsonian Center for Astrophysics,
60 Garden St, Cambridge, MA 02193, USA}

\author[0000-0002-2503-3269]{Helle Bakke}
\author[0000-0003-2088-028X]{Luc Rouppe van der Voort}
\affil{Rosseland Centre for Solar Physics, University of Oslo, P.O. Box 1029 Blindern, N-0315 Oslo, Norway}
\affil{Institute of Theoretical Astrophysics, University of Oslo, P.O. Box 1029 Blindern, N-0315 Oslo, Norway}

\author[0000-0002-8370-952X]{Bart De Pontieu}
\affil{Lockheed Martin Solar \& Astrophysics Laboratory,
3251 Hanover St, Palo Alto, CA 94304, USA}
\affil{Rosseland Centre for Solar Physics, University of Oslo, P.O. Box 1029 Blindern, N-0315 Oslo, Norway}
\affil{Institute of Theoretical Astrophysics, University of Oslo, P.O. Box 1029 Blindern, N-0315 Oslo, Norway}


\begin{abstract}
High resolution spectral observations of the lower solar atmosphere (chromosphere and transition region) during coronal heating events, in combination with predictions from models of impulsively heated loops, provide powerful diagnostics of the properties of the heating in active region cores. Here we analyze the first coordinated observations of such events with the {\em Interface Region Imaging Spectrograph} (\iris) and the CHROMospheric Imaging Spectrometer (CHROMIS), at the Swedish 1-m Solar Telescope (\sst), which provided extremely high spatial resolution and revealed chromospheric brightenings with spatial dimensions down to $\sim 150$~km. We use machine learning methods ($k$-means clustering) and find significant coherence in the spatial and temporal properties of the chromospheric spectra, suggesting, in turn, coherence in the spatial and temporal distribution of the coronal heating. The comparison of \iris\ and CHROMIS spectra with simulations suggest that both non-thermal electrons with low energy (low-energy cutoff $\sim 5$~keV) and direct heating in the corona transported by thermal conduction contribute to the heating of the low atmosphere. This is consistent with growing evidence that non-thermal electrons are not uncommon in small heating events (nano- to micro-flares), and that their properties can be constrained by chromospheric and transition region spectral observations.
\end{abstract}
 

\section{Introduction}
\label{introduction}
The understanding of the physical processes converting magnetic energy into thermal energy and powering the solar outer atmosphere represents one of the main open issues in solar physics \citep[e.g.,][]{Klimchuk2006,Reale2014,Testa_Reale_2022arXiv}. The heating is generally predicted to be impulsive \citep[e.g.,][]{Klimchuk2015}, and on small spatial scales (below the current resolution capabilities).  Direct observational diagnostics of coronal heating are therefore often difficult to obtain. 

Tracers of coronal heating can sometimes be more evident in observations of the lower solar atmosphere -- transition region (TR) and chromosphere -- which efficiently radiates energy excesses, rather than in the highly conductive corona, where the signatures of heating release are easily washed out. 
Furthermore, while the coronal emission in a pixel is due to contributions from a generally complex three dimensional system of loops, the TR is relatively unencumbered by contamination of other material along the line of sight. It is therefore easier to detect and study single heating events in the TR, and derive constraints on the coronal event properties. For these reasons, the moss (the bright TR of high pressure loops; e.g., \citealt{Peres94,Fletcher1999,Berger1999}) is well suited to investigate the properties of coronal heating in the core of ARs \citep[e.g.,][]{Martens00,Antiochos2003,Testa2013,Testa2014,Testa2020}. 

The temporal variability of the moss has been studied in imaging and spectral observations, and its relatively constant emission has often been attributed to steady heating of AR cores \citep[e.g.,][]{Antiochos2003,Brooks2009,Tripathi2010}. However, high spatial ($\sim 0.3$\arcsec) and temporal resolution EUV imaging data with the High-resolution Coronal Imager (Hi-C; \citealt{Kobayashi2014}) sounding rocket, have provided evidence of TR brightenings on short timescales (down to $\sim 15$~s) at the footpoints of transient hot loops \citep{Testa2013}. Follow-up TR spectral observations at high spatial, temporal, and spectral resolution with the Interface Region Imaging Spectrograph (\iris, \citealt{DePontieu14}), together with 1D hydrodynamic RADYN models of nanoflare heated loops -- including non-local thermodynamic equilibrium (non-LTE), and heating by beams of non-thermal electrons (NTE) as well as in-situ thermal heating--, provided powerful diagnostics of the properties of coronal heating and mechanisms of energy transport \citep{Testa2014,Testa2020,Cho2023}.
\cite{Polito2018} and \cite{Testa2020} discussed the \iris\ diagnostics of the coronal heating properties from spectral observations of moss brightenings.  
\cite{Bakke2022} recently extended those investigations to ground-based observations, deriving additional diagnostics of the heating properties from lower chromospheric emission.

\begin{figure*}[!ht]
	\centering
    \includegraphics[width=18cm]{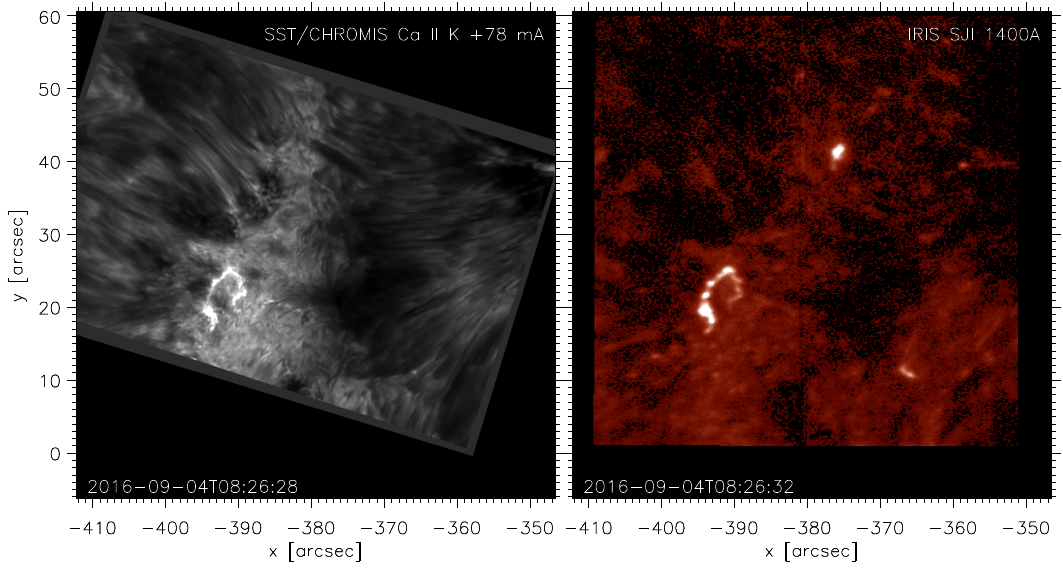} 
 	\caption{Chromospheric and TR observations in AR 12585 with \sst/CHROMIS and \iris\ on 2016-09-04 around 08:26UT. {\em Left}: \sst/CHROMIS image at the center wavelength of \caii\ (at 3934\AA); {\em Right:} \iris\ slit-jaw image (SJI) in the 1400\AA\ band. The bright S-shaped region corresponds to brightenings at the footpoints of hot loops undergoing short-lived heating, as illustrated in the context coronal images presented in Figure~\ref{fig:fig_ar_img_bands}. 
	}
	\label{fig:fig_sst_iris}
\end{figure*}

In this paper we analyzed coordinated chromospheric AR observations, with the Swedish 1-m Solar Telescope (\sst; \citealt{Scharmer2003a,Scharmer2003b}), and with \iris, as well as coronal observations with \sdo/AIA, for a small heating event in an AR. 
The very high resolution chromospheric observations spectra from \sst\ provide new constraints to the models of small heating events.  A $k$-means analysis of these \sst\ spectra allows us to detect coherence in the spatial and temporal distribution of the chromospheric emission, and, in turn, to speculate on the coherence of the spatial and temporal distribution of the heating properties.
\iris\ spectral observations provide insights into the chromospheric and TR properties. 
In Section~\ref{sec:obs} we describe the selected data, and their analysis and comparison with expectations from numerical simulations is presented in Section~\ref{sec:results}. In Section~\ref{sec:discussion} we summarize and discuss our findings, and draw our conclusions.

\section{Observations and Data Analysis}
\label{sec:obs}

The detection of small (nano- to microflare) coronal heating events in coordinated \iris\ and \sst\ observations of active regions is one of the goals of the multi-year campaign we have conducted since the \iris\ launch in 2013 \citep{Rouppe2020}. In fact, \sst\ and \iris\ provide very useful complementary high-resolution observations of the photosphere and chromosphere that can greatly enhance the scientific impact of either separate instrument.

\begin{figure*}[!ht]
	\centering
    \hspace{-0.5cm}
    \includegraphics[width=17.5cm]{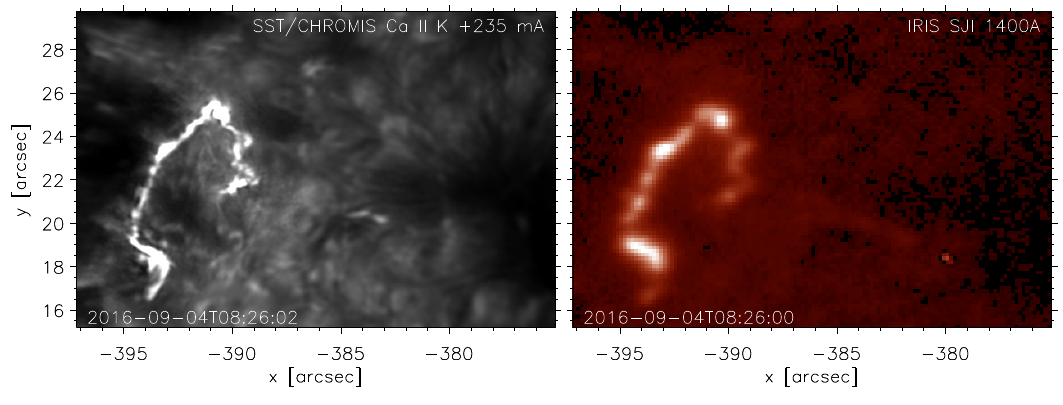}\vspace{-0.2cm}
    
\hspace{-1cm}
	\includegraphics[width=18cm]{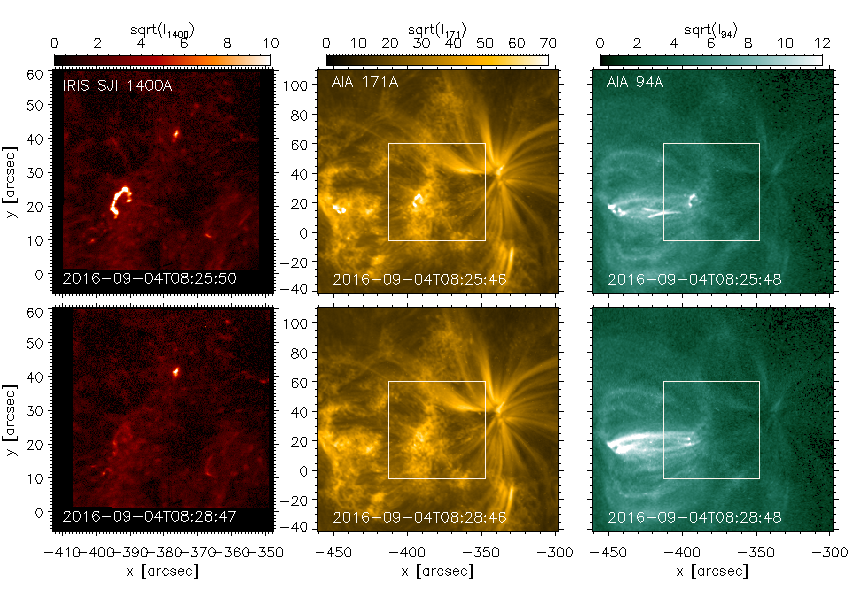}\hspace{-1cm}	
	\caption{\sst, \iris, and \sdo\ coordinated observations of AR~12585. In the top panels we plot zoomed in versions of the images of Figure~\ref{fig:fig_sst_iris} showing emission in the \caii\ K core ({\em left}) and \iris\ 1400\AA\ SJI ({\em right}). The bottom panels show the TR and coronal emission, for two times -- one at the peak of the loop footpoint chromospheric emission ({\em top}), and one 3~min later ({\em bottom}) when the overlying hot coronal loops are bright. In particular we show, from left to right: the TR emission in the \iris\ SJI 1400\AA\ passband; the upper TR emission in the AIA 171\AA\ narrowband, dominated by $\sim 1$~MK emission from \feix; the coronal emission in the AIA 94\AA\ narrowband (which has a cooler, $\sim 1$~MK, component, and a hot $\gtrsim 4$~MK component). For the AIA emission we show a larger f.o.v.\ than the corresponding \iris\ observations.  We mark with white boxes the f.o.v.\ of the \iris\ SJI.
     }
	\label{fig:fig_ar_img_bands}
\end{figure*}

Here we focus on one of the first such events observed during the \sst\ and \iris\ coordinated campaign. In particular, we selected the observations of AR~12585 carried out on 2016-09-04 (Figure~\ref{fig:fig_sst_iris} and Figure~\ref{fig:fig_ar_img_bands} show images of this event observed by \sst, \iris, and \sdo). This dataset is part of the database of publicly available \sst\ and \iris\ coaligned datasets described in \cite{Rouppe2020}, and some analysis of it has been presented by \cite{Rouppe2017}.

The \sst\ telescope includes dual Fabry-P{\'e}rot filtergraph systems capable of fast wavelength sampling of spectral lines: the CRisp Imaging SpectroPolarimeter (CRISP; \citealt{Scharmer2008}), and the CHROMospheric Imaging Spectrometer (CHROMIS; \citealt{Scharmer2017}), with field of view of approximately $1\arcmin \times 1\arcmin$.
CRISP has a plate scale of 0\farcs058 per pixel and the \sst\ diffraction limit is 0\farcs14 at the wavelength of the H$\alpha$ line (calculated as $\lambda/D$ with $\lambda=6563$\AA\ and $D=0.97$~m the diameter of the \sst\ aperture). The CHROMIS instrument (installed in 2016) has a plate scale of 0\farcs038 per pixel and the diffraction limit is 0\farcs08 at the wavelength of the \caii\ K line ($\sim 3933.7$\AA). The data we analyze here were acquired during the first CHROMIS campaign \citep{Rouppe2017,Vissers2019}. 
Here we mainly focus on the analysis of the chromospheric \caii\ K data.
The \caii~K line was sampled at 21 wavelength positions, between $\pm$100~\kms\ Doppler offset from line center. The line was sampled with 6~\kms\ steps (or 78~m\AA) between $\pm$54~\kms\ and somewhat coarser in the rest of the wavelength range. In addition, a continuum position was sampled at 4000~\AA.
The CRISP instrument was running a program sampling the \Halpha\ line at 15 line positions between $\pm$68~\kms, and \caii~8542\AA\ at 21 line positions between $\pm61$~\kms.
The data was processed following the CRISPRED reduction pipeline \citep{2015A&A...573A..40D} and an early version of the CHROMISRED pipeline 
\citep[now both are incorporated into SSTRED,][]{2021A&A...653A..68L}. 
Seeing-induced deformations were corrected for using Multi-Object Multi-Frame Blind Deconvolution \citep[MOMFBD;][]{2005SoPh..228..191V}  image restoration. 
The CHROMIS timeseries has a cadence of $\sim 22$~s, the CRISP timeseries has a cadence of 20~s.
The wavelength calibration of the SST/CHROMIS observations has been performed by using an average profile over a quiet area just prior to the observations, and deriving the wavelength within the Ca II line where the intensity is at a minimum value. This reference wavelength is given the value (3933.6841\AA) of the wavelength of the 
\caii\ K line minimum in the FTS atlas \citep{Neckel1999}.

\begin{figure*}[!ht]
	\centering
    \includegraphics[height=6cm]{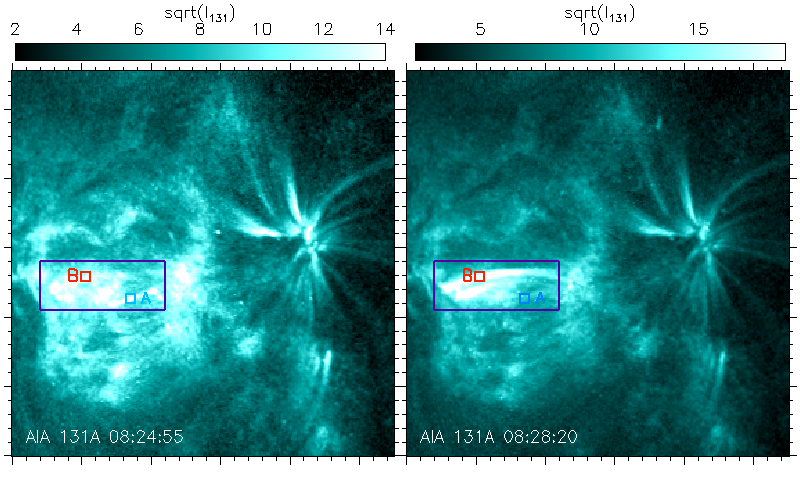}
	\includegraphics[height=6cm]{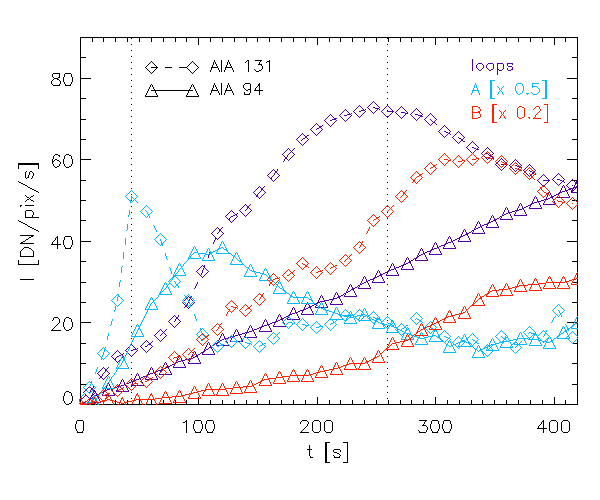}
	\caption{Evolution of the hot coronal emission in the AR core during the heating event. The left two panels show the coronal emission in the \sdo/AIA 131\AA\ passband (the 131\AA\ emission in the transient core loops is dominated by hot \fexxi\ emission), and for the same f.o.v.\ of the other AIA passbands shown in Figure~\ref{fig:fig_ar_img_bands}, at two times (close to the beginning of the heating event and a few minutes later). Three locations are marked in these images to identify interesting coronal features: a short hot loop brightening early in the event ({\em A}), a location in a longer loop at the footpoints of which the chromospheric and TR brightenings are observed as shown in Figures~\ref{fig:fig_sst_iris} and \ref{fig:fig_ar_img_bands} ({\em B}), and a larger region (purple rectangle) including most of the dynamic hot loops anchored in the ribbon shown in the previous two figures. The right panel shows lightcurves ($t=0$ corresponds to 2016-09-04T08:24:00) in the hot AIA passbands at 131\AA\ (diamonds), and 94\AA\ (triangles), for the three regions marked in the left panels. For the {\em A} and {\em B} locations the intensities are averaged over $5 \times 5$ pixel regions (i.e., $\sim 3$\arcsec$ \times 3$\arcsec). All lightcurves are subtracted of their minimum value in the considered temporal window}. The vertical dotted lines mark the times of the two images shown in the left panels.
	\label{fig:fig_aia_lc}
\end{figure*}

We note that particularly the CHROMIS observations have exceptionally high spatial resolution ($\sim 100$~km), which can constrain significantly better the spatial distribution of the heating events. 

\iris\ observes chromospheric and TR emission in UV at high spatial and temporal resolution with both high-resolution spectroscopy and slit-jaw imaging (SJI).  
The \iris\ observations we analyze here are medium dense 16-step rasters (OBSID 3625503135), with exposure time of 0.5~s, and raster steps of 0\farcs35, covering a field-of-view of about 5\arcsec\ $\times$ 60\arcsec\ (in about 21~s).
Slit-jaw images were recorded in the SJI 1400~\AA\ (dominated by \siiv\ lines), 1330~\AA\ (dominated by \cii\ lines), and 2796~\AA\ (\mgii~k core) channels at a temporal cadence of about 10~s. 
We use \iris\ calibrated level 2 data, which have been processed for dark current, flat field, and geometrical corrections \citep{DePontieu14}. 
The \sst\ and \iris\ observations were aligned through cross-correlation of image pairs in the \caii~K wing and SJI 2796\AA\ \citep[see][for more details]{Rouppe2020}. 
The alignment between \caii~K and SJI 2796\AA\ appears to be accurate down to the \iris\ SJI pixel size (0\farcs166). The short exposure time of the \iris\ program (0.5~s) resulted in a high noise level in the SJI images, in particular in SJI 1400\AA\ and SJI 1330\AA.  The fiducial marks that are added to the \iris\ spectrograph slit are difficult to identify in individual SJI images and even impossible to identify in SJI 1400\AA. This makes the alignment between the SJI channels challenging as the fiducial marks are the basis for accurate cross alignment. To improve on the SJI alignment as part of the standard \iris\ level 2 data processing, we summed over many SJI exposures to reduce the noise level and improve the visibility of the fiducial marks. It is clear that this procedure improved the alignment but it cannot be excluded that there remains a residual misalignment on the order of one or two pixels between SJI 2796\AA\ and SJI 1400\AA.

We also use coordinated imaging observations of coronal emission, taken with the Atmospheric Imaging Assembly (AIA; \citealt{Lemen2012}) onboard the Solar Dynamics Observatory (\sdo; \citealt{Pesnell2012}). The AIA datasets are characterized by 0\farcs6 pixels, and 12~s cadence, and observe the TR and corona across a broad temperature range \citep{Boerner2012,Boerner2014}. 
We used the AIA datacubes coordinated with and coaligned to the \iris\ datasets, which are distributed from the \iris\ search data page\footnote{https://iris.lmsal.com/search/}. These AIA time series appear to be affected by unusual and significant (i.e., several AIA pixels) instabilities in the spatial pointing. We therefore refined the coalignment, by using a cross-correlation routine (tr\_get\_disp).
We use AIA coronal data to investigate the spatial and temporal properties of the coronal emission.

\section{Results} 
\label{sec:results}

AR~12585 appeared at the East limb around August 30 2016, and it was characterized by sustained coronal activity around GOES (Geostationary Operational Environmental Satellites\footnote{https://www.swpc.noaa.gov/products/goes-x-ray-flux, https://www.ngdc.noaa.gov/stp/satellite/goes/index.html}) B level. An inspection of the timeseries of the hot emission in the \sdo/AIA channels (especially the 94\AA\ and the 131\AA\ channels, which include emission from \fexviii\ and \fexxi\ lines respectively) shows hot ($\gtrsim 5$~MK) dynamic loops in the AR core, although the X-ray emission in the GOES X-ray passbands did not exceed C level for most of the disk passage (from September 1st onward). 

The observations we analyze here captured one of these heating events associated with hot coronal emission. In Figure~\ref{fig:fig_ar_img_bands} we show the chromospheric, TR, and coronal emission in the AR at two times of this event: early on, when the energy is most likely impulsively released and the chromospheric and TR emission at the loops' footpoints is bright, and a few minutes later, when the coronal plasma in the overlying loops is sufficiently hot and dense to emit brightly in the AIA 94\AA\ passband. 

In the initial impulsive phase of the heating, the chromosphere and TR present rapid and intense brightenings at the footpoints of the heated loops, similar to flare ribbons, as observed by \sst\ and \iris\ and shown in Figure~\ref{fig:fig_sst_iris} and Figure~\ref{fig:fig_ar_img_bands} where almost simultaneous images of the chromospheric and TR emission are presented. 

\subsection{Coronal Properties and Evolution} \label{sec:corona}
The heating event appears to involve many coronal structures, including the longer ($\sim 40$~Mm in length) coronal structures overlying the bright ribbons we focus on, as well as shorter ($\sim 20$~Mm in length) loops overlying a ribbon brightening about a minute earlier (see center row of Figure~\ref{fig:fig_ar_img_bands}, and Figure~\ref{fig:fig_aia_lc}).
In Figure~\ref{fig:fig_aia_lc} we show lightcurves for the hot emission observed by AIA in the 94\AA\ and 131\AA\ channels. The lightcurves are obtained by subtracting the background value in the same spatial location just before the event, to highlight the evolution of the hot emission. The observed hot emission shows the typical behavior observed in other microflares (see e.g., \citealt{Testa_Reale_2020}), with a rapid increase in the \fexxi\ emission (131\AA) in the early phases of the event followed by a cooling phase in which the 131\AA\ emission decreases while the cooler \fexviii\ 94\AA\ emission increases. An approximate temperature diagnostic from the 131\AA/94\AA\ ratio (see e.g., \citealt{Testa_Reale_2020}) suggests peak coronal temperatures around 8--10~MK, analogous to what typically found in similar transient events in AR cores \citep[e.g.,][]{Reale2019a,Testa2020,Testa_Reale_2020}.

\begin{figure*}[!ht]
	\centering
    \includegraphics[width=15cm]{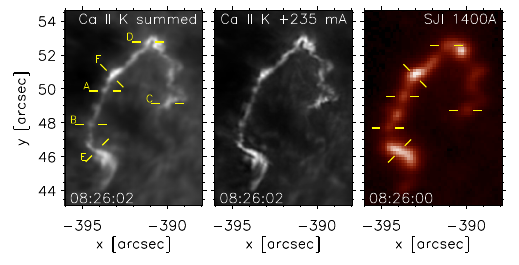}     \includegraphics[width=15cm]{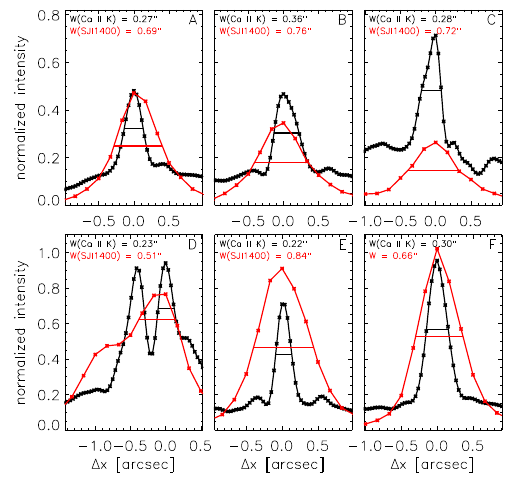}
    \caption{Cross-section widths of brightenings in the flare ribbon from CHROMIS and \iris\ observations. The top left panel shows an image integrated over the full observed width of the \caii~K line ($\pm100$~\kms), the top middle panel shows a \caii~K red wing image at around the peak of most redshifted flare profiles. The top right panel shows the corresponding \iris\ SJI 1400\AA\ image. All images are scaled linearly between their respective minima and maxima. The yellow lines mark the endpoints of the the paths A--F along which intensity profiles were extracted and are shown in the bottom panels. In the bottom panels, the black line is the intensity profile measured in \caii~K +0.235\AA, the red line in SJI 1400\AA. The full-width-half-maxima are marked with horizontal lines and their values are given in each panel.   }
    \label{fig:fwhm}
\end{figure*}

\begin{figure*}[ht]
	\centering
	\includegraphics[width=18cm]{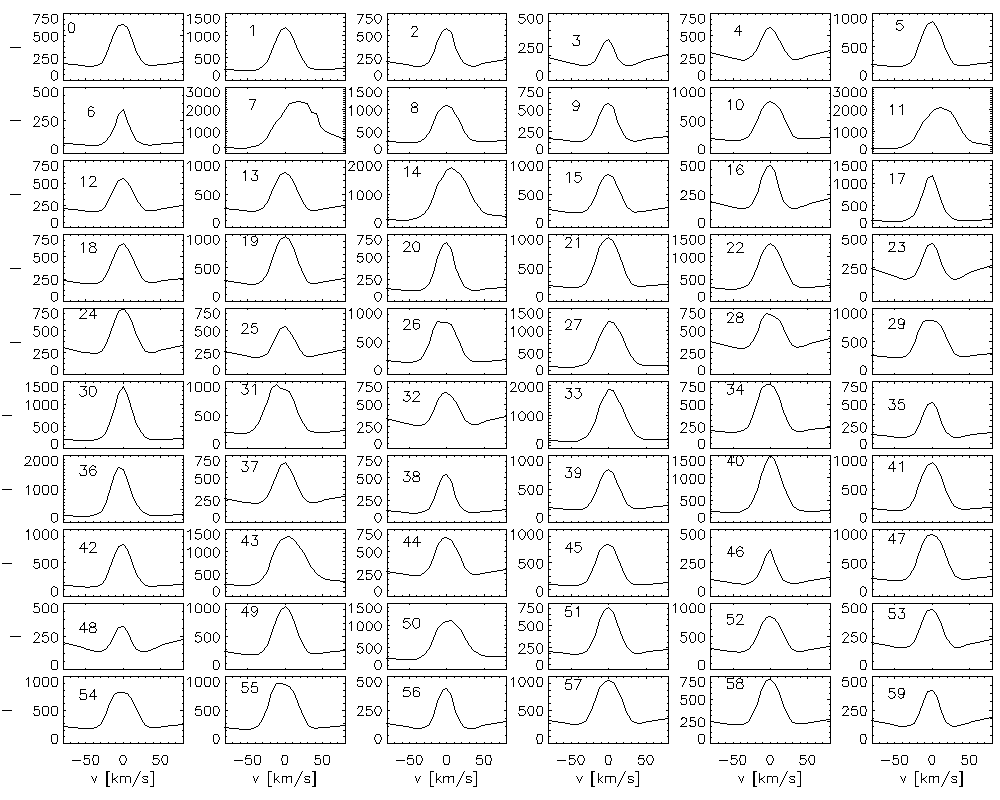}	
	\caption{Representative Spectral Profiles (RSP) obtained from $k$-means analysis, using 60 clusters (see text for a discussion), of the timeseries of \sst/CHROMIS \caii\ K spectra.}
	\label{fig:fig_rsp}
\end{figure*}

\subsection{Morphology of the Ribbon} \label{sec:ribbon}
The comparison of the two \sst\ and \iris\ emission maps shown in Figures~\ref{fig:fig_sst_iris} and \ref{fig:fig_ar_img_bands} clearly illustrates the differences in morphology in different atmospheric layers, as well as (particularly in Figure~\ref{fig:fig_ar_img_bands}) the extremely high spatial resolution attained in this \sst\ dataset by the CHROMIS instrument.
Figure~\ref{fig:fwhm} shows measurements of the cross-section widths of the ribbon brightenings. 
There are small spatial offsets of about 0\farcs2 between corresponding brightenings in \caii~K and SJI 1400\AA. 
Given differences in formation heights and the geometry of the observed coronal loops, spatial offsets could be expected and they might provide valuable information on the 3D morphology of the loop system.
We can expect a small offset between the different diagnostics due to differences in formation height, but we are cautious about drawing firm conclusions since we cannot exclude that the offset might be due to a large extent to errors in the alignment between the different passbands. Furthermore, the ribbons evolve very fast and the \sst\ and \iris\ diagnostics are not recorded strictly simultaneously so some of the apparent offsets could be attributed to temporal evolution. 

We measured the cross-section widths as full-width-half-maxima for which the half maxima were determined from the difference between the peak intensity and the highest intensity of the two neighboring minima.
For these six representative examples, the widths in \caii~K vary between 0\farcs22 and 0\farcs36.
The widths in SJI 1400\AA\ are more than two times larger and vary between 0\farcs51 and 0\farcs84. 
We note that the narrowest width of 0\farcs51 corresponds to three \iris\ pixels. 
In feature D, the cross-section profile of the \caii~K image has two narrow peaks while the corresponding feature in SJI 1400\AA\ is much wider and fuzzier and does not show sub-structure.
The profiles A--F are shown for red wing \caii~K at +235~m\AA\ (+18~\kms) which is the Doppler offset of the peak for most of the redshifted ribbon profiles. The widths of cross-section profiles from \caii~K images summed over the full observed spectral widths are only slightly wider. 
We note how, closer to the two ends of the ribbon (i.e., S of $y \sim 45$, or W of $x \sim -390$, such as e.g., location C at the N-W end), the relative intensity of the CHROMIS \caii\ emission with respect to the \iris\ SJI 1400\AA\ emission is much larger than for most other locations, whereas in other locations such as e.g., E, the TR emission is significantly enhanced with respect to the lower chromospheric emission. As we will discuss more later (see Section~\ref{sec:discussion}), these ratios provide clues to the heating properties and transport mechanisms (e.g., harder non-thermal electrons will likely cause stronger enhancements in the deeper atmosphere, i.e., lower chromosphere, than in the TR).

\begin{figure*}[ht]
	\centering
    \includegraphics[height=5.76cm]{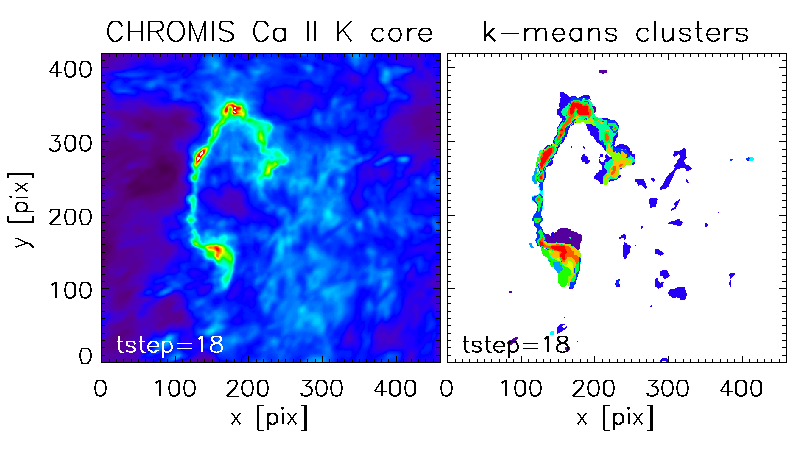}
    \includegraphics[height=5.76cm]{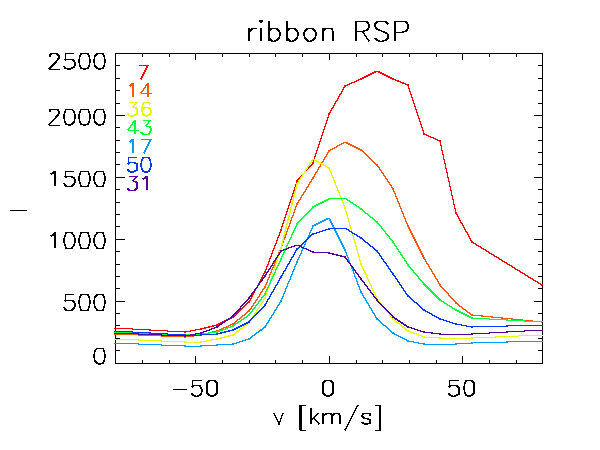}	
	\caption{Example of machine learning analysis of \sst/CHROMIS chromospheric spectral observations of variability at the footpoints of hot loops. We show results of $k$-means clustering analysis applied to a subset of the \sst/CHROMIS \caii\ K spectral data. We selected the f.o.v.\ shown here and 14 timesteps covering the heating event shown in Figure~\ref{fig:fig_sst_iris}, and a few timesteps prior to it. The $k$-means clustering algorithm finds spectral profiles representative of the observed spectra, and it groups the observed spectra according to their spectral properties (in this example we find 60 clusters are sufficient to group the observed spectra). Here we show, for one time step during the brightenings (the intensity in the \caii\ K line core is shown in the left panel), the spatial distribution (middle panel) of the few clusters occurring in the footpoint regions, and the corresponding representative spectral profiles (RSP; right panel; the $k$-means analysis is run on the whole time series and it produces a single set of RSPs). The RSPs present a variety of properties including broad blueshifted (cluster 31; purple) and redshifted profiles (e.g., clusters 7, 14,  43, 50).}
	\label{fig:fig_chromis_kmeans1}
\end{figure*}

\subsection{k-means Analysis of \sst\ Chromospheric Spectra} \label{sec:kmeans}
To investigate the spatial and temporal evolution of the spectral properties of the chromospheric brightenings we applied $k$-means clustering analysis \citep[e.g.,][]{Panos2018,Bose2019} to the \sst/CHROMIS \caii\ K spectral data. The $k$-means clustering algorithm finds spectral profiles representative of the observed spectra, and it groups the observed spectra according to their spectral properties. The  grouping of the large number of observed spectra in a limited number of clusters, each characterized by a representative spectral profile (RSP), allows us to model more easily the variety of observed profiles, and, in turn, to efficiently derive the spatial distribution and temporal evolution of the coronal heating properties.
We used the $k$-means clustering algorithm \citep{everitt1972cluster}, which partitions data into $k$ pre-defined clusters that each have a cluster center which is the average of the data points within that cluster. The clusters are then improved through an iterative process where data points are assigned to a cluster such that the Euclidean distance between the points and the cluster center is minimal. For every iteration, new cluster centers are calculated from the clusters in the previous iteration until the cluster centers do not change and convergence is reached. A limitation to this method is that the resulting clusters depend significantly on the initial selection of clusters centers. An option is to select the cluster centers at random, but then the initial cluster centers can be close to each other and more iterations are needed in order to reach convergence. We used the $k$-means\texttt{++} initialization method \citep{arthur2007k}, which, before defining new cluster centers, draws the cluster centers randomly from the dataset such that they are as far away as possible from the previous centers.

\begin{figure*}[ht]
	\centering
	\includegraphics[width=16cm]{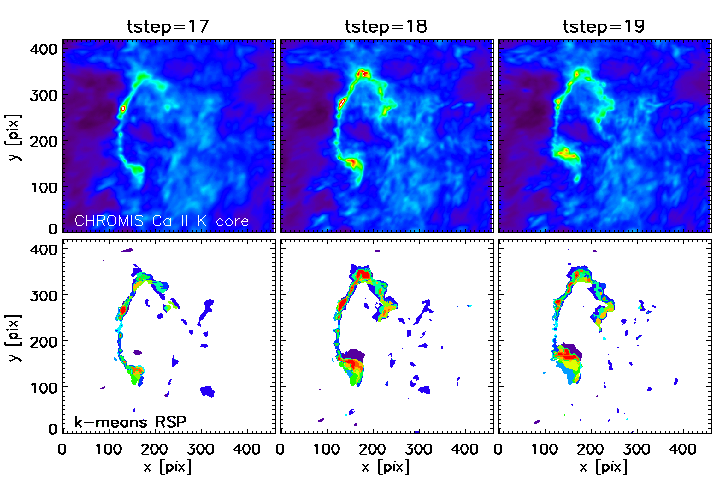}\vspace{-0.1cm}	
	\includegraphics[width=16cm]{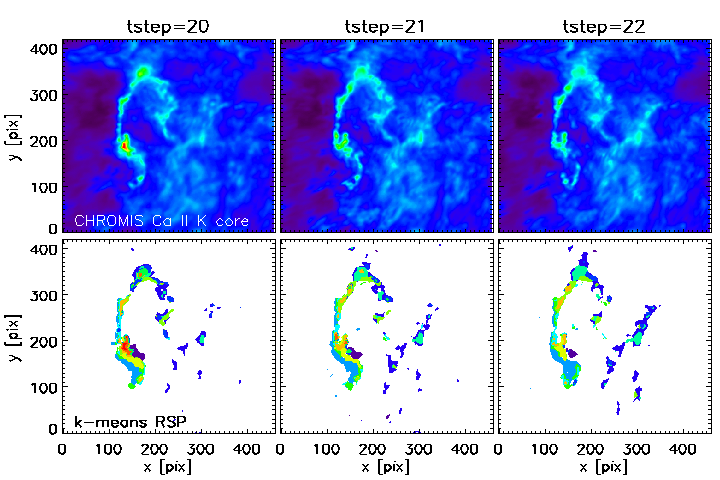}
	\caption{Results of the $k$-means clustering analysis allow to investigate the spatial and temporal evolution of the spectral properties of the chromospheric brightenings.
    Here we show, for six consecutive time steps (about 22~s apart) during the brightenings (the intensity in the \caii\ K line core is shown in the top panels for each time step), the spatial distribution (bottom panels for each time step) of the few clusters occurring in the footpoint regions, corresponding to the representative spectral profiles shown in Figure~\ref{fig:fig_chromis_kmeans1} (right panel).}
	\label{fig:fig_chromis_kmeans2}
\end{figure*}

In order to determine the optimal number of clusters, we used the elbow technique (see e.g., \citealt{Panos2018} for a discussion), and we found that 60 clusters are sufficient to characterize the selected subset of \sst/CHROMIS spectral observations.
The 60 representative spectral profiles (RSP) we obtained from this $k$-means analysis are plotted in Figure~\ref{fig:fig_rsp}. 
In Figure~\ref{fig:fig_chromis_kmeans1} we show the spatial distribution of the RSPs in the region of the ribbon, and the corresponding RSPs, which show that the \caii~K line observed in the ribbon varies from blueshifted, and generally narrower, profiles to redshifted, and typically broader, profiles. While the observed blueshifts are generally mild ($\lesssim 10$~km~s$^{-1}$), some redshifts reach large values (up to $\sim 25$~km~s$^{-1}$). The largest blueshifts are mostly observed in the southern portion of the ribbon. The spatial distribution of the RSPs shows a spatial coherence of the observed spectral profiles, with spatial clusters of several sizes from a few pixels ($\gtrsim$0\farcs1) to $\sim 50$~pixels ($\sim 2$\arcsec), in one spatial dimension.

We investigated how the spatial distribution of the \caii\ K spectral profiles is evolving during the event, as shown in Figure~\ref{fig:fig_chromis_kmeans2}, where we plot the maps of RSP distribution for six consecutive timesteps of the CHROMIS observations covering most of the ribbon evolution. These plots show that the spatial coherence of the spectral profiles is observed for all timesteps, and they reveal several additional interesting features. The highest Doppler shifts are not present in the initial \caii~K profiles (time step 17), but they appear in the second timestep when higher intensities are observed in the ribbon.  In fact, the highest \caii~K intensity regions are generally characterized by large and broad redshifted profiles (RSP 7 and 14, red and orange respectively). Blueshifted profiles are observed only in lower intensity regions of the ribbon, and only in the southern portion of the ribbon.  In this southern region of the ribbon (around coordinates [160, 170] in time step 18; see Figure~\ref{fig:fig_chromis_kmeans1}) there is a patch of broad blueshifted profiles (RSP 31 of Figures~\ref{fig:fig_rsp} and ~\ref{fig:fig_chromis_kmeans1}), where the spectra rapidly shift to broad redshifted profiles in the next time step (19) when the \caii~K intensities in that region increase significantly. We note that for similar events observed with \iris\ \citep[e.g.,][]{Testa2014,Testa2020,Cho2023} the observed chromospheric and TR spectral profiles significantly change on short timescales of seconds. In light of this, the relatively modest cadence of these CHROMIS data might not provide us with a comprehensive view of the spectral evolution at the loops footpoints.
The temporal evolution of the \caii~K spectra in the spatial locations where the highest intensities are observed appear to generally follow a progression from broad profiles with largest redshifts at the peak intensity to progressively less redhifted profiles (i.e., red to orange/yellow/green, in terms of the RSPs shown in Figure~\ref{fig:fig_rsp}).
Fig.~\ref{fig:fig_chromis_kmeans2} shows that, as the event progresses, the ribbon, especially the southern portion, advances mostly in the north direction. As new locations in the lower atmosphere get brighter they tend to be characterized initially by blueshifted and narrow \caii~K profiles, and then evolve to brighter, broader and more redshifted profiles. This observed evolution of the spectral profiles could be due to the change in the plasma conditions in a magnetic strand, where increasing density and temperature as a consequence of the heating cause a different response to the heating properties, and/or could point to different heating properties during the events, with harder non-thermal electron distributions in newly reconnected lines evolving in time to softer distributions.

\begin{figure*}[ht]
	\centering
	\includegraphics[width=\textwidth]{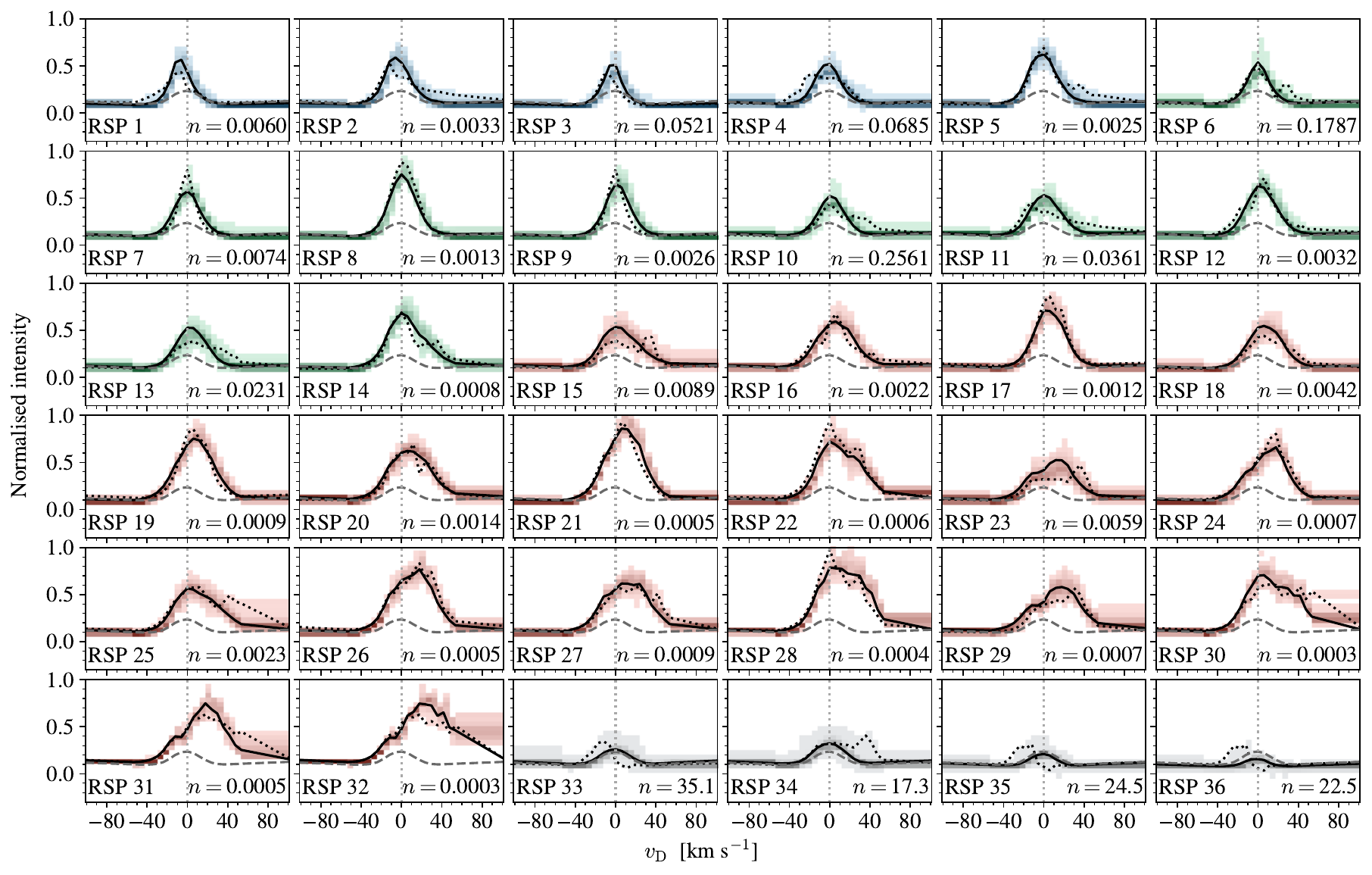}
	\caption{Thirty-six RSPs from the $k$-means clustering of the \caii\ K line profiles from a reduced data set focused on the ribbon (see also Figure~\ref{fig:fig_cluster_maps}). Each panel shows the RSPs (solid black), the average profile in the reduced data (dashed gray), and the \caii\ K profile that have the largest Euclidean distance to the RSP (dotted black). All profiles have been normalised from the lowest to the highest intensity of the reduced data set. The panels include the density distribution of all profiles within a cluster, where darker color indicates higher density. The color of the density distributions in RSPs 1--32 also represent the Doppler shift of the RSPs to indicate blueshifts (blue), redshifts (red), and weak to no shifts (green). RSPs 33--36 are clusters of profiles outside the flare ribbon, and the gray shading is the density distribution of all profiles in the clusters. 
	    RSPs 1--32 are sorted by the Doppler shift of the Gaussian profiles that were fitted to the RSPs, and RSPs 33--36 are placed at the end of the sorting. $n$ represents the number of profiles in a cluster as a percentage of the total amount of profiles in the reduced data set ($\sim 2.25 \times 10^7$ profiles).}
	\label{fig:fig_rsp_ribbon}
\end{figure*}

In the above $k$-means clustering analysis applied to the \sst\ field of view which includes both ribbon and non-ribbon emission, only a few clusters describe the ribbon spectra (see Figure~\ref{fig:fig_chromis_kmeans1}).  Therefore, in order to refine our analysis of the specific spectral features of the flare profiles, 
we run a $k$-means analysis on a subset mostly comprised of spectral profiles from the flare ribbon. We used a reduced data set that covers 550 pixels in $x$ and 500 pixels in $y$ centered on the flare ribbon, including the 82 time steps. To find the profiles covering the flare, we sampled all profiles with maximum intensity above 1565 in arbitrary intensity units (see contours in Figure~\ref{fig:fig_cluster_maps}). This value was found by exploring different thresholds and see how adjusting it affected the clustering of flare profiles. We also included profiles from outside the ribbon, creating a total sample of 94\,823 profiles, where 62\% are flare profiles and 38\% are profiles sampled in the surroundings but outside the flare. We set the number of clusters to 36 in order to capture more details of the profiles, as well as to make sure that specific features end up in less populated clusters. The $k$-means algorithm was applied to the training data set, and the output was used to predict the closest RSPs for each profile in this reduced data set.

\begin{figure*}[ht]
	\centering
	\includegraphics[width=\textwidth]{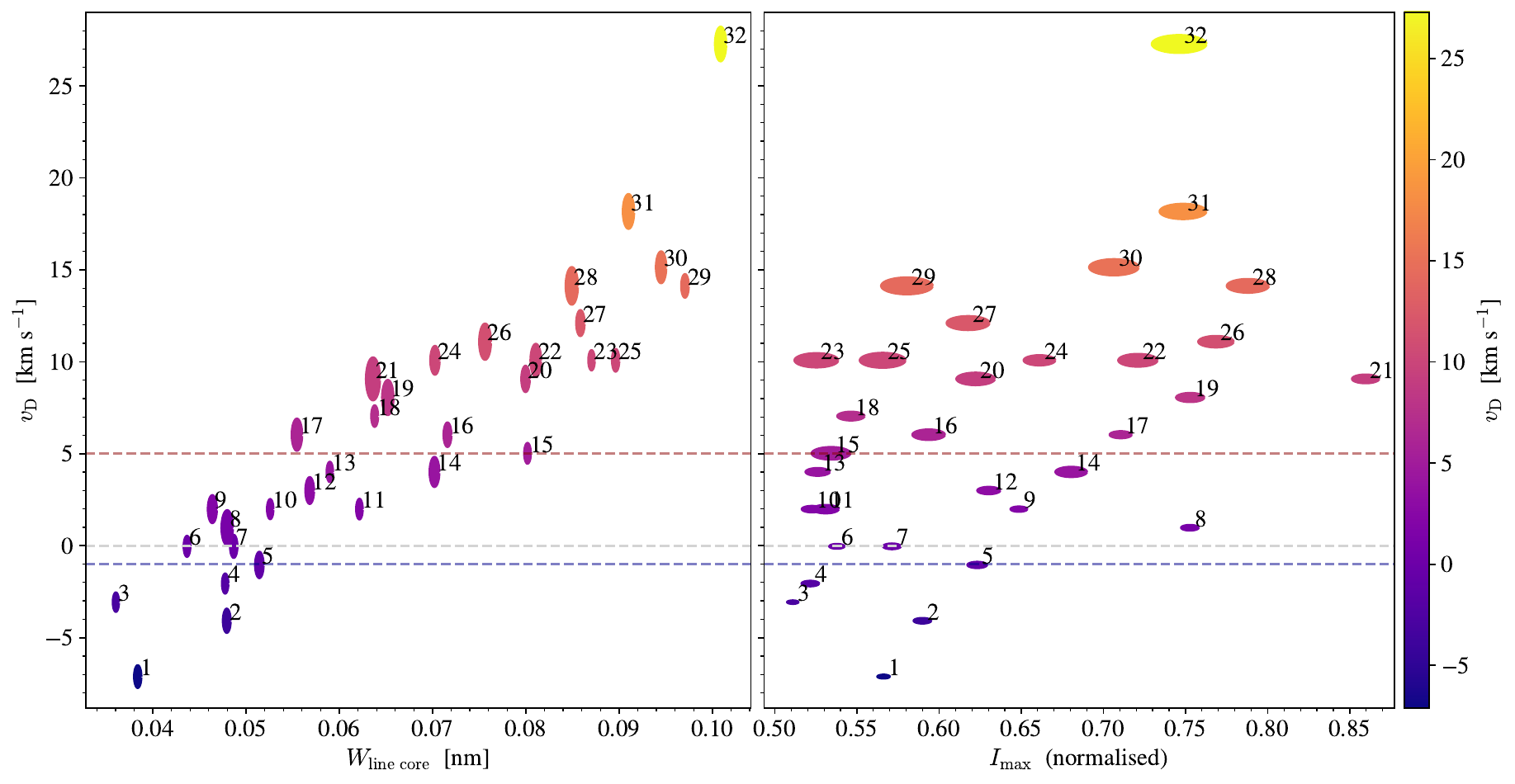}
	\caption{Scatter plots between the Doppler shift and RSP line core widths (left), and Doppler shift and maximum intensity of the RSPs (right). The points are marked with their respective RSP index (see Figure~\ref{fig:fig_rsp_ribbon}), and colored based on their Doppler shift. The vertical radii of the ellipses in the left figure reflect the magnitude of the maximum intensity of the RSPs, while the horizontal radii in the right figure represent the RSP line core widths. The red dashed line at $+5$~\kms\ and blue dashed line at $-1$~\kms\ indicate the boundaries for the RSP sorting.}
	\label{fig:fig_scatter_rsp}
\end{figure*}

Figure~\ref{fig:fig_rsp_ribbon} shows an overview of the 36 clusters. We applied a Gaussian fit to the RSPs in order to calculate their Doppler shifts and used these values to order the clusters according to their Doppler shift. 
The number of redshifted RSPs is significantly higher than the blueshifted RSPs. In addition, the Doppler shifts to the red generally have higher values. Because of this, we set two different thresholds on the RSPs characterized by redshifts and blueshifts: $+5$~\kms\ and $-1$~\kms, respectively (see red and blue dashed lines in Figure~\ref{fig:fig_scatter_rsp}).
Following these thresholds, Figure~\ref{fig:fig_rsp_ribbon} shows that RSPs 1--5 are shifted to the blue, RSPs 6--14 have weak to no Doppler shifts, and RSPs 15--32 are shifted to the red. RSPs 33--36 are large clusters containing the profiles from outside the ribbon, and their respective RSPs have 0~\kms\ Doppler shift. We note that the profiles in the latter clusters are only used as a reference and are not included in the rest of the analysis, hence the corresponding RSPs are placed at the end of the sorting. 
The figure shows that the $k$-means clustering is able to identify and properly cluster specific features. This is seen from the density distributions, showing that the majority of profiles in each cluster are centered around or close to the RSPs. 

\begin{figure*}[ht]
	\centering
	\includegraphics[width=0.8\textwidth]{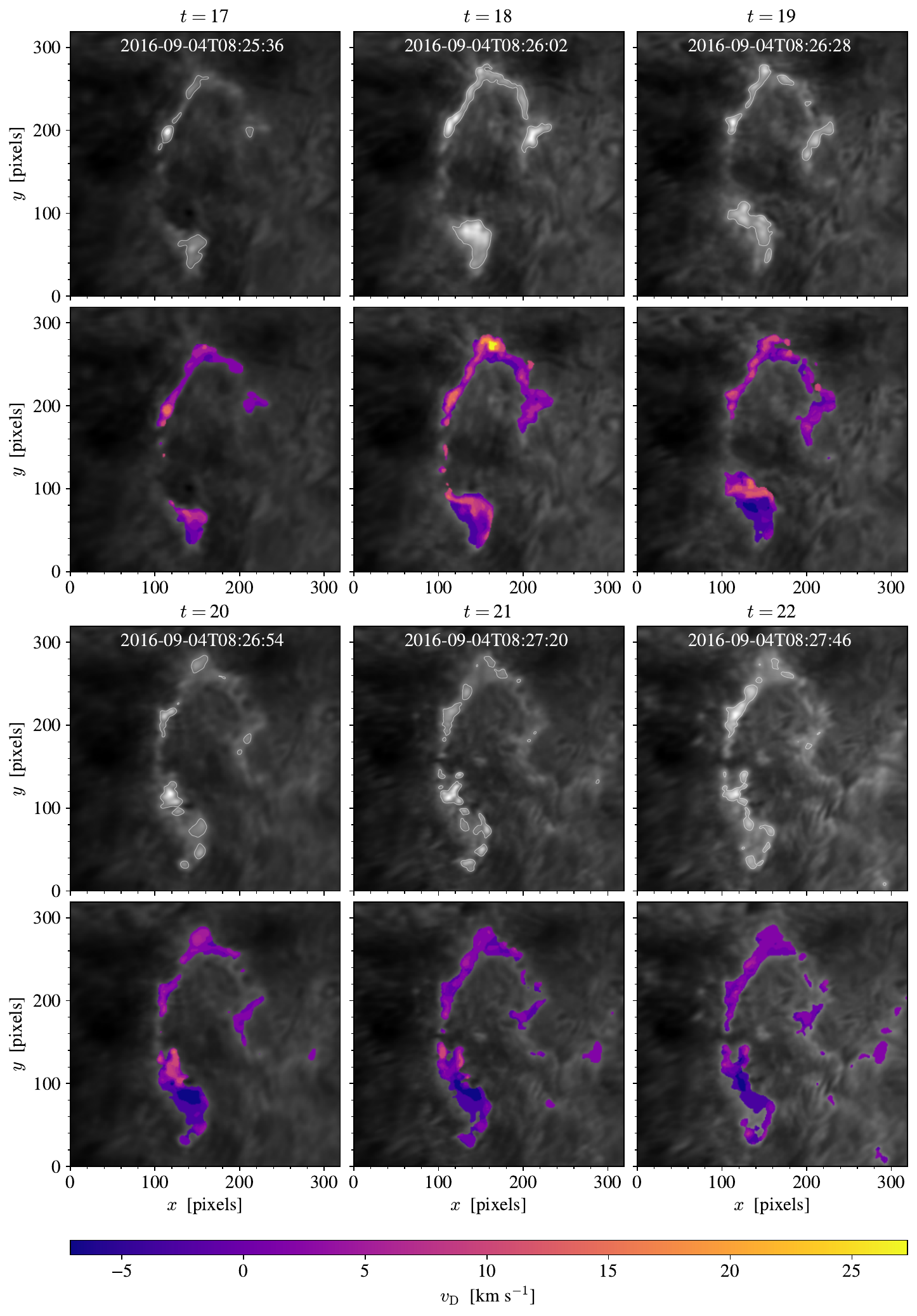}
	\caption{Intensity maps of the \sst/CHROMIS \caii\ K line core for six consecutive time steps during the flare. The contours in the top panels of each time step show the threshold (contours) used for the sampling of the $k$-means training data, for the analysis focused on the ribbon spectra. In the bottom panels of each time step we overplot, on the intensity maps, the map of Doppler shift of the RSPs (clusters 1--32 shown in Figure~\ref{fig:fig_rsp_ribbon}). 
    The f.o.v.\  has been reduced to mostly cover the ribbon.}
	\label{fig:fig_cluster_maps}
\end{figure*}

Figure~\ref{fig:fig_scatter_rsp} shows two scatter plots between the Doppler shift and RSP line core widths (left) and the Doppler shift and maximum intensity of the RSPs (right). The color of the ellipses follows the respective Doppler shift of the RSPs. The vertical radii of the ellipses in the left panel is proportional to the maximum intensity of the RSPs, while the horizontal radii of the ellipses in the right panel is proportional to the width of the RSP line cores. The left panel shows that there is a correlation between the Doppler shift and the line core width, where larger Doppler shifts mostly result in broadening of the line cores. This is also clearly seen from the horizontal radii of the ellipses in the right panel. This panel also shows that the maximum intensity of the RSPs is not strongly correlated to the Doppler shift.

\begin{figure*}[ht]
	\centering
	\includegraphics[width=\textwidth]{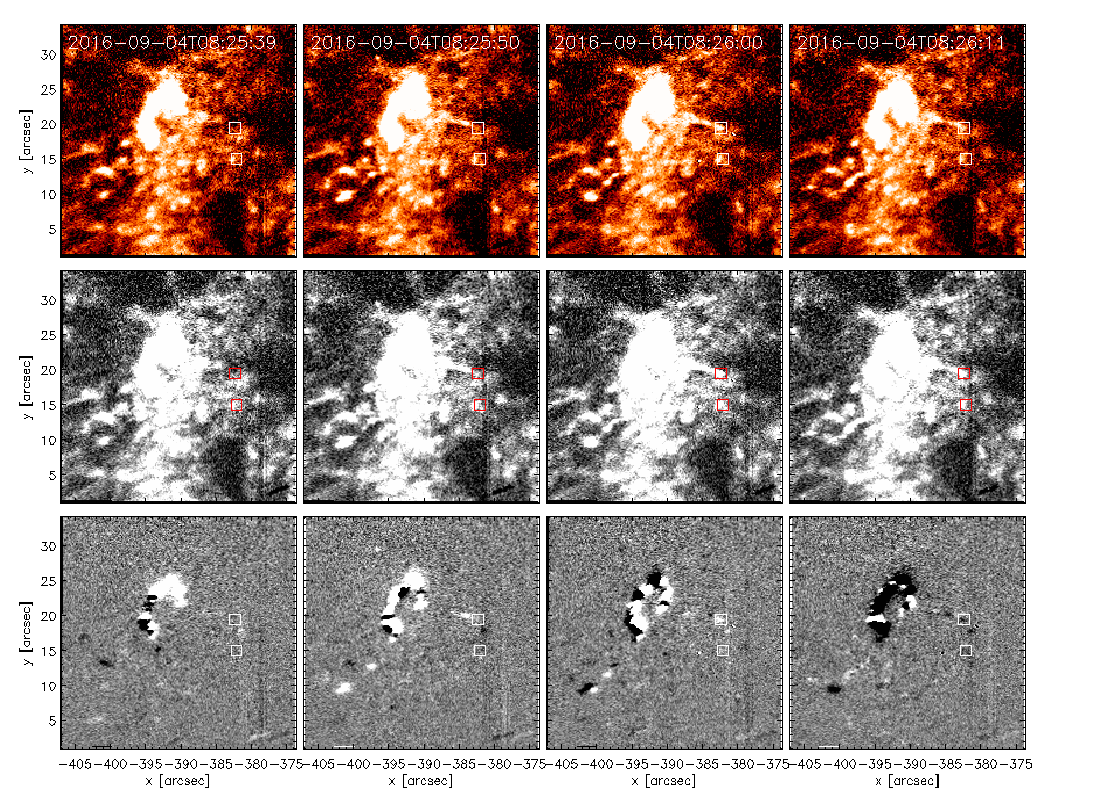}
	\caption{\iris\ SJI time series around a time when part of the ribbon is under the \iris\ slit. From top to bottom: \iris\ 1400\AA\ SJI emission (top); base difference (i.e., the emission subtracted of the image at a reference time, which we chose to be 2016-09-04T08:19:35; middle); and running difference (i.e., the difference between the current SJI image and the SJI image from the immediately preceding time step; bottom). The two squares indicate the position for which we show the \iris\ and CHROMIS \caii\ spectra in Figure~\ref{fig:iris_spec} (for better contrast we use a red color for the boxes in the middle panels). The emission of the ribbon is quite weak at those locations, motivating the saturated color scales we used in the plots of this figure to highlight their emission as much as possible.  }
	\label{fig:iris_sji_diff}
\end{figure*}

The Doppler shift of the flare profiles is further explored in Figure~\ref{fig:fig_cluster_maps}, showing intensity maps of the \sst/CHROMIS \caii\ K line core at six consecutive time steps during the flare. While the top panels of each time step are overplotted with contours marking the threshold for the $k$-means training sampling, the bottom panels show the Doppler shift of every profile in RSPs 1--32. The results are similar to those found in Figure~\ref{fig:fig_chromis_kmeans2}, where blueshifted profiles are exclusively found in low-intensity areas in the lower part of the ribbon and are not present until $t = 19$. The most redshifted profiles (RSPs 31 and 32) only occur at $t = 18$ in the top part of the ribbon. Even though the profiles in these clusters exhibit the strongest redshifts, they are generally not the most intense profiles found in the flare. While there indeed are a few profiles in these clusters with high intensity (see density distribution in Figure~\ref{fig:fig_rsp_ribbon}), RSP 21 contains profiles with significantly higher intensities but much lower Doppler shifts (approximately $+9$~\kms). This cluster is also a clear outlier in the right panel in Figure~\ref{fig:fig_scatter_rsp}. We generally see that the regions with high \caii\ K intensity are characterized by redshifted profiles, but we note that there are still lower-intensity regions with profiles that are significantly redshifted and broadened (for example RSPs 23, 25, 27, and 29) as well as high-intensity regions with profiles that are less redshifted and not as broad (for example RSPs 8, 17, 19, and 21). 

\begin{figure*}[ht]
	\centering
	\includegraphics[width=\textwidth]{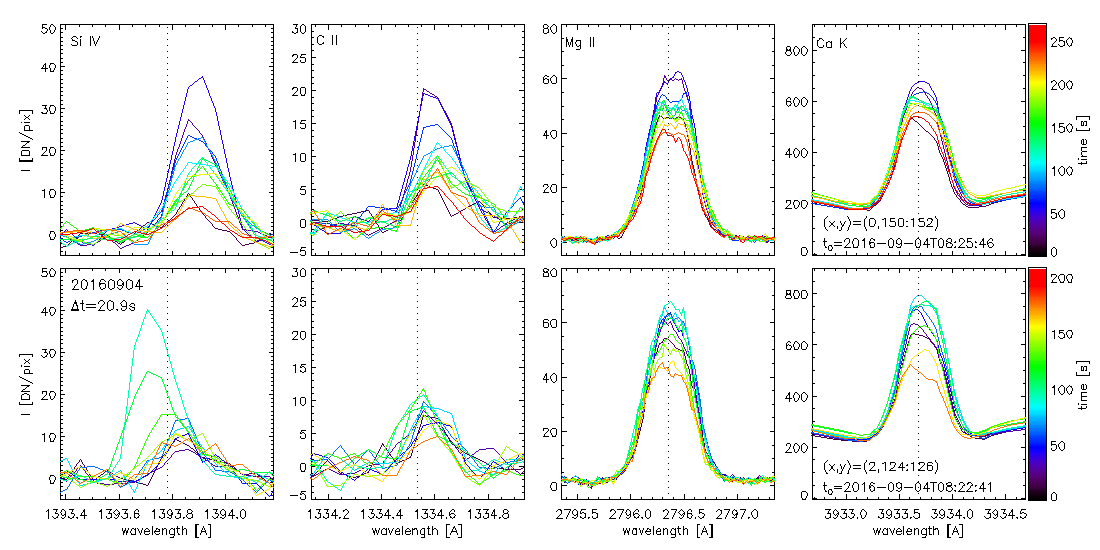}
	\caption{Temporal evolution of the spectral emission in \iris\ TR and chromospheric lines and in the \sst/CHROMIS \caii\ K line (from left to right: \siiv\ 1393\AA, \cii\ 1334\AA, \mgii\ 2796\AA, \caii\ 3934\AA) for two locations (one in each row) where faint part of the ribbons is under the \iris\ slit. For the CHROMIS \caii\ spectra we use 3933.684\AA\ as a reference wavelength (see section~\ref{sec:obs} for details). The top and bottom row correspond respectively to the northernmost and southernmost boxes of Figure~\ref{fig:iris_sji_diff}. Note that the two time series have different timescales (see colorbars to the right) and start time ($t_0$ is noted in the right plot).}
	\label{fig:iris_spec}
\end{figure*}

\subsection{\iris\ and \sst\ Spectra} \label{sec:spectra}
\iris\ can provide additional chromospheric and TR diagnostics. Although the brightest portion of the ribbon-like region is unfortunately not under the \iris\ slit, the slit crosses a weaker portion of the ribbon, where we can analyze the \iris\ and CHROMIS spectral profiles. In Figure~\ref{fig:iris_sji_diff} we highlight this weak ribbon region by showing base difference and running difference \iris\ SJI time series: for the base difference we subtract the initial emission (we chose 2016-09-04T08:19:35 as reference time), whereas for the running difference we subtract the image of the immediately preceding time step. 
These figures show the presence of a ``tail" of the ribbon extending from (x,y) $\approx (-387",22")$ to (x,y) $\approx (-382",18")$, where the latter is in the \iris\ spectrograph f.o.v. 
Another location under the \iris\ slit, around (x,y)$\sim (-382",17")$, is also undergoing short-lived brightenings a couple of minutes earlier, and appears to be associated with the early brightening of the shorter coronal loops (see left panel of Figure~\ref{fig:fig_aia_lc}). 
In Figure~\ref{fig:iris_spec} we show the timeseries of the spectral profiles in some of the strongest \iris\ TR and chromospheric lines, and in the CHROMIS \caii~K, for these two locations (note that we average the \iris\ spectra over 3 \iris\ pixels along the y axis to improve the signal-to-noise).
The earlier brightening (bottom row of Figure~\ref{fig:iris_spec}, and southernmost square marked in Figure~\ref{fig:iris_sji_diff}) shows: (1) a significant increase in the \siiv\ TR emission lasting about a minute and characterized by a blueshift of $\sim 20$\kms\ (relative to the pre-event spectrum), and a broad (multi-peaked) profile with a small red tail; (2) a very modest increase in \cii\ emission; (3) modest increase in chromospheric \mgii\ emission, which is mostly characterized by limited central reversal, and a blue peak slightly more pronounced than the red peak; (4) the CHROMIS \caii\ K profiles are similar to the profiles observed in the regions of the main ribbon with weaker emission, i.e., they are single peaked, narrow, and with modest Doppler shift (e.g., RSPs 17, 36 of Figures~\ref{fig:fig_rsp} and \ref{fig:fig_chromis_kmeans1}, or RSPs 7--13 of Figure~\ref{fig:fig_rsp_ribbon} and \ref{fig:fig_scatter_rsp}).
The second brightening, in the weak ``tail" of the ribbon, (top row of Figure~\ref{fig:iris_spec}, and northernmost square marked in Figure~\ref{fig:iris_sji_diff}), is slightly longer lasting (about a couple of minutes), and it shows: (1) \siiv\ emission without significant Doppler shift with respect to the pre-event profile (but redshifted in absolute terms), (2) \cii\ emission with relative increase, with respect to the pre-event profile, comparable to the \siiv\ emission; (3) the chromospheric \mgii\ and \caii\ emission are similar to the other brightening, with largely single-peaked profiles and CHROMIS \caii\ K profiles similar to the weaker ribbon regions.
For both brightenings there is no significant \mgii\ triplet emission detected.
In the following subsection (Section~\ref{sec:sims}) we will compare the observed profiles with predictions from 1D RADYN models of impulsively heated loops. 

\subsection{Comparison with Numerical Simulations} \label{sec:sims}

Numerical simulations can provide a great deal of insight when interpreting observations. In previous work \citep{Polito2018, Testa2020, Bakke2022} we used 1D flare simulations to investigate the atmospheric response to heating by nanoflares. Several models were carried out using the RADYN numerical code \citep{1992ApJ...397L..59C, 1995ApJ...440L..29C, 1997ApJ...481..500C, 2015TESS....130207A}, 
which models plasma magnetically confined in a 1D loop structure, by solving the equation of charge conservation and the level population rate equations.
RADYN also solves the non-local thermodynamic equilibrium (non-LTE) radiation transport for H, He, and \caii\, which is necessary when modeling chromospheric emission. The RADYN version described in \cite{2015TESS....130207A} 
includes the effect of non-thermal electrons, using the Fokker-Planck equations, and the electron energy distribution is assumed to follow a power-law. Details on the RADYN numerical code and the simulations of nanoflare-heated loops are discussed in \cite{Polito2018}. 
 In the following we provide a brief description of the models.

\begin{figure*}[ht]
	\centering
	\includegraphics[width=\textwidth]{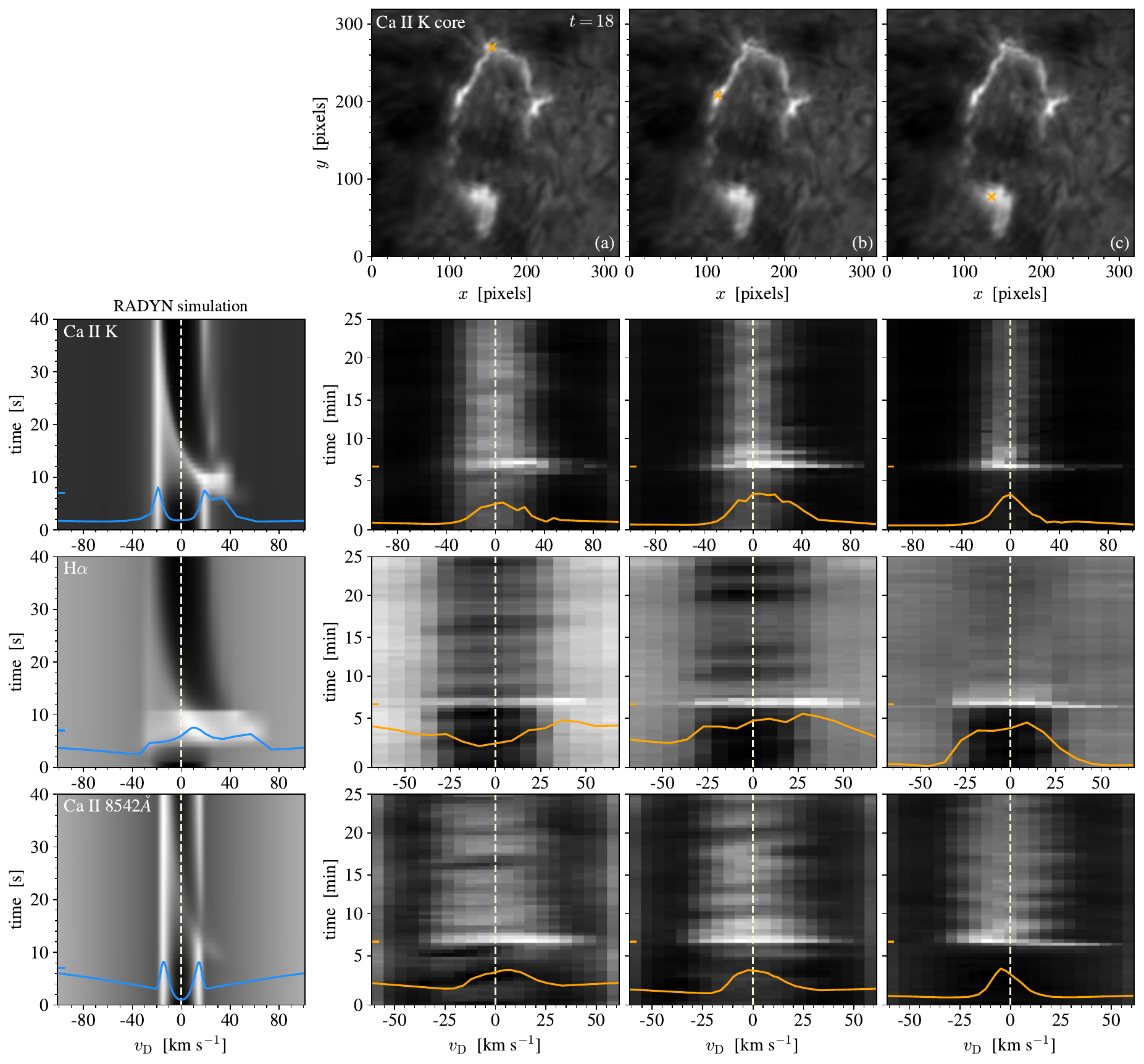}
	\caption{Spectral evolution of \caii\ K, H$\alpha$, and \caii\ 8542\AA\ in a RADYN simulation of a nanoflare heated loop and from \sst\ observations at different locations in the ribbon. The top panels show intensity maps of the \sst/CHROMIS \caii\ K line core at $t = 18$, where the ribbon locations at which the spectra are taken are marked as orange crosses. Each spectral evolution panel (the three bottom rows) shows the line profile at a time step during the flare phase. For the RADYN simulation, the time step is chosen at 7~s while the line profiles from the observations are chosen at 6.8~min (which is at $t = 18$). These time steps are marked along the $y$-axis. We note that the scaling of the $x$-axes varies between the observational datasets.}
	\label{fig:fig_radyn_obs}
\end{figure*}

The RADYN simulations of nanoflare-heated loops investigate a broad parameter space including different nanoflare energies, loop-top temperatures, and half-loop lengths. The simulations also include different heating models, such as thermal conduction, electron beam heating with varying low-energy cutoff values $E_\mathrm{C}$, as well as hybrid models of both thermal conduction and electron beams. In \cite{Bakke2022}, we focused our efforts on selected models with transport by non-thermal electrons, where the chromospheric lines were synthesized using the RH1.5D radiative transfer code \citep{2001ApJ...557..389U, 2015A&A...574A...3P}.
By comparing the \caii\ K, H$\alpha$, and \caii\ 8542\AA\ synthetic spectra from these models to the flare profiles from the observations, we find that the RADYN model most capable of reproducing the observed spectra is the model with 15~Mm half-loop length and initial loop-top temperature of 1~MK. This particular model has a low-energy cutoff value $E_\mathrm{C} = 5$~keV (representing an electron beam with low-energy electrons in the distribution), a spectral index $\delta = 7$ for the power-law energy distribution, and total energy $E = 6 \cdot 10^{24}$~erg deposited in the loop. 
Figure~\ref{fig:fig_radyn_obs} shows the time evolution of \caii\ K, H$\alpha,$ and \caii\ 8542\AA\ from the RADYN simulation and at three different locations in the flare ribbon, where the latter is given in the top row panels. The locations were chosen in order to show the time evolution of the profiles in a region of strong redshifts (panel a), a region of high intensity (panel b), and a region of blueshifts (panel c). 
We note that there is a discrepancy between the flare time scales of the RADYN simulation and the observation. In the RADYN model, the flare lasts for 10~s, while in the observation the flare lasts for minutes. The RADYN model was originally meant to simulate short-lived (10--30~s) footpoint brightenings in TR moss as observed with \iris\ \citep{Testa2013, Testa2014, Testa2020,Cho2023}. 
Even though the observed flare is a sequence of many short-lived brightenings, it becomes difficult to make a temporal comparison between the profiles because the time between each observed frame is about half the total simulation time. 
The comparisons made in this work are therefore focused on the shape and features of the profiles rather than the time scales. 

The shape of the RADYN \caii\ K and \caii\ 8542\AA\ profiles does not resemble that of the observations, especially since the absorption feature of the synthetic spectral lines is not seen in the observed flare profiles.
However, the strong redshift of the \caii\ K profile is consistent with the observation at the peak of the flare (around 6.8~min) at the the locations marked in panels (a) and (b) of Figure~\ref{fig:fig_radyn_obs}. The largest Doppler shift of the \caii\ K line observed in the northern part of the ribbon (panel a) is comparable to that of the RADYN simulation, where the lines from both the observation and simulation have components that are redshifted to a value between 30 and 35~\kms.

H$\alpha$ is the synthetic profile from RADYN that is most similar to the observational profiles. In particular, the H$\alpha$ profile from the location marked in panel (b) has an almost identical shape to the one from RADYN, although with a slightly smaller range of velocity on the red side of the line profile. 
The evolution of the profiles is also similar, where the initial profiles are in absorption until the flare phase during which both profiles are in emission and characterized by a redshifted component with multiple peaks. In RADYN, the H$\alpha$ profile  eventually returns back to being in absorption during the post-heating phase (after 10~s). The heating mechanisms of the actual flare are more complicated, and after the peak of the flare there is still ongoing heating that continuously changes the shape, width, and intensity of the profiles over time. However, at the northern (column a) and central (column b) locations of the ribbon, the profiles seem to slowly revert back into absorption, which is consistent with the simulation. The location in the southern region of the ribbon (column c) shows less variation of the H$\alpha$ intensity and profile features after 8~min compared to the other locations.

The comparison of the \iris\ spectral properties during the brightenings, with the expectations from RADYN models, can guide us in the interpretation of the observations and how they constrain the heating properties. As discussed at length in previous work \citep{Testa2014,Polito2018,Testa2020}, the grid of RADYN simulations of impulsively heated loops that we have carried out, provide diagnostics of the presence of non-thermal particles and their properties. In particular, blueshifts in the \siiv\ profiles, or \mgii\ triplet emission, are signatures of accelerated particles. 
The \iris\ spectral properties of the earlier brightening are overall reminiscent of model H1 (which is a hybrid model, with half-length of 15~Mm, where half of the energy is transported by thermal conduction and half goes into nonthermal particles) of \cite{Testa2020}, in particular, the multicomponent nature of the \siiv\ profile with a blueshifted peak, the slightly higher blue peak in the \mgii\ profiles, and the lack of significant \mgii\ triplet emission. We can speculate that the \iris\ profiles, and modest intensities of the brightenings,  suggest that the heating is due to a mix of direct heating with thermal conduction and non-thermal particles, characterized by low energy (likely around 5~keV) and total energy and flux likely smaller than in our existing simulations. In fact, as we discussed in \cite{Polito2018}, impulsive heating with non-thermal particles can produce \siiv\ blueshifts for lower energy cutoff values when the total energy is reduced (their Figure~15). As discussed earlier in this section, such a model produces H$\alpha$ profiles also similar to the observed ones.
For the second brightening, the spectral profiles do not clearly point to the presence of non-thermal particles, but also for this case some of the observed spectral properties are qualitatively similar to the models with low energy non-thermal electrons ($E_\mathrm{C}$ $\sim 5$~keV), in particular in terms of the lack of significant Doppler shift in \siiv\ and of significant \mgii\ triplet emission (see also Fig.~2 of \citealt{Cho2023}). 

There are a few caveats to keep in mind when using the RADYN simulations of impulsively heated loops to interpret the observations we are analyzing here. 
First, these simulations do not reproduce the observed \mgii\ and \caii\ (and often also the \cii) profiles, in particularly failing to reproduce the often observed single-peak profiles, as discussed above (and in previous works, such e.g., \citealt{Testa2020}).
The mismatch between modeled and observed chromospheric profiles occurs also in quiescent plasma (e.g., \citealt{Carlsson2015,Hansteen2023}), and suggests that the background atmosphere might be a non-negligible cause of the problem.
Furthermore, these simulations, as described earlier in this section, were developed to match short-lived brightenings typically observed to last less than one minute, and therefore might not be an ideal comparison for the observations in this paper which have longer duration ($\sim 2$--4~min) footpoint brightenings.
Also, the observations we analyzed have about 21~s cadence (and only $\sim 0.5$~s exposure time) therefore the temporal sampling is such that it might have missed a crucial initial phase of the brightening(s), including possibly the peak of the emission.

\section{Discussion and conclusions}
\label{sec:discussion}

In this paper we have analyzed coordinated imaging and spectroscopic observations of a small heating event in the core of an active region, observed with \iris, \sdo/AIA, and the CHROMIS instrument at \sst. The atmospheric response to the heating includes the initial brightening at the footpoints of the coronal loops, visible in the chromospheric and TR emission in \iris, \sst, and AIA data, and the subsequent brightenings of the coronal loops, first in the hot AIA 131\AA\ (\fexxi) emission, followed by the 94\AA\ (\fexviii) and other cooler channels during the cooling phase. 

The morphology of the hot coronal loops, with sets of transient loops crossing at a significant angle, is reminiscent of what is generally observed in other similar heating events in AR cores \citep[e.g.,][]{Testa2013,Testa2014,Reale2019a,Testa2020,Testa_Reale_2020}, suggesting such events might be driven by large-scale photospheric motions or large-scale magnetic flux emergence (see also \citealt{Asgari2019}).
The coronal emission points to high temperatures up to $\sim 10$~MK, analogous to other small heating events in AR cores (see e.g., \citealt{Brosius2014,Ishikawa2017,Ishikawa2019,Reale2019a,Reale2019b,Testa2020,Testa_Reale_2020,Cooper2021}). During the overall heating event, a first set of footpoint brightenings are observed (around 8:24UT) including one location under the \iris\ slit (bottom row of Figure~\ref{fig:iris_spec}, and southernmost location marked in Figure~\ref{fig:iris_sji_diff}) and the footpoints of a set of short loops (sampled by location A of Figure~\ref{fig:fig_aia_lc}). The brightest chromospheric/TR brightenings are observed about two minutes later at the footpoints of a longer set of loops (sampled by location B of Figure~\ref{fig:fig_aia_lc}), and are well observed by CHROMIS and the \iris\ SJI (see e.g., Figure~\ref{fig:fig_sst_iris}), while only a weaker tail of this ribbon is observed under the \iris\ slit (top row of Figure~\ref{fig:iris_spec}, and northernmost location marked in Figure~\ref{fig:iris_sji_diff}). 

The simultaneous \iris\ and CHROMIS observations provide complementary diagnostics of the heating event. This CHROMIS dataset is in particular characterized by exceptional spatial resolution down to $\sim 100$~km.
The spatial analysis of the \sst\ observations of the ribbon indicates spatial scales of $\sim 150$--200~km (measured FWHM $\sim$0\farcs2-0\farcs3) for the chromospheric emission. There is a small observed spatial offset ($\sim$0\farcs2) between the \sst\ chromospheric emission and the TR \iris\ SJI emission in a direction compatible with the observed geometry, i.e., with the coronal loops extending in the E direction, and therefore the TR also extending to the E of the lower atmospheric emission observed by \sst, although we cannot completely rule out small misalignment errors (see discussion in Section~\ref{sec:ribbon}). 
The relative intensity of the TR (\iris\ SJI 1400\AA) to lower chromospheric emission (observed with CHROMIS) is observed to vary over the ribbon, in particular with relatively weaker TR emission close to the two ends of the ribbon (i.e., S of $y \sim 45$, or W of $x \sim -390$, such as e.g., location C of Figure~\ref{fig:fwhm}), which suggests (see e.g., Fig.~5, and discussions in \citealt{Testa2020}) that in those locations the heating and energy transport mechanisms might less efficiently heat the TR, indicating for instance less direct heating and thermal conduction, and/or harder non-thermal electrons, compared to other ribbon locations.

CHROMIS provides simultaneous spectral data in every spatial pixel, and therefore a trove of constraints for the spatial and temporal properties of the heating. In order to more efficiently exploit this information, we applied machine learning methods to the CHROMIS \caii\ K data, and in particular $k$-means clustering analyses which can efficiently sort the observed spectra into groups with similar spectral properties (Representative Spectral Profiles, RSP). This analysis reveals that the chromospheric emission is characterized by spatial and temporal coherence, in turn suggesting spatial and temporal coherence of the heating properties. In particular, in most ribbon locations the \caii\ spectra rapidly evolves from initial narrower profiles with modest Doppler shift toward redshifted and broader profiles (see Figures ~\ref{fig:fig_chromis_kmeans2} and \ref{fig:fig_cluster_maps}), as the ribbon evolves and moves, mostly northward. This connection between spectral evolution and ribbon motion puts tight constraints on models. The observed spectral evolution can likely be ascribed to either an evolution of the heating properties (e.g., harder non-thermal electron distributions in newly reconnected lines) and/or the evolution of the atmosphere as it gets denser and hotter, in response to the heating.  A small fraction of \caii\ spectra is blueshifted, and those blueshifted profiles are mostly concentrated in the southern region of the bright ribbon.  The blueshifted profiles are generally narrower, and with lower intensities, than the redshifted profiles (see Figure~\ref{fig:fig_scatter_rsp}).
The very interesting correlations between \caii\ Doppler shift, line width, and intensity provide valuable constraints for models.
If the Doppler shifts of the \caii\ spectral profiles are linked to plasma bulk velocities in the lower chromosphere, the correlations between the Doppler shift and line width could indicate a relationship between the plasma flows and the causes of line broadening (e.g., opacity broadening, superposition of plasma flows along the line-of-sight). 

For a couple of low brightness ribbon locations, we obtained \iris\ and CHROMIS spectra that could be used for a more constraining comparison with predictions from RADYN models of impulsively heated loops. For these locations, the \caii\ K profiles are similar to what is observed in the low intensity regions of the main ribbon, and in \iris\ no significant \mgii\ triplet emission is observed (which would have been a clear signature of high energy non-thermal particles), and the \mgii\ spectra show mostly single-peaked profiles. The \iris\ \siiv\ and \cii\ emission shows slightly different properties in the two locations, with the earlier brightening showing blueshifted \siiv\ spectra and small \cii\ enhancement, and the later brightening characterized by no significant Doppler shift change during the heating event and comparable increase of emission in \siiv\ and \cii. 

The comparison of these observed spectral properties with the prediction from models generally suggests the presence of low energy non-thermal particles (with low-energy cutoff $\sim 5-10$~keV), likely accompanying direct heating in the corona transported by thermal conduction. We note however that the grid of models we are using \citep{Polito2018,Testa2020,Bakke2022} was developed to reproduce observed brightenings with shorter duration ($< 1$~min) than the ones observed here. Also, as we discussed in detail in Section~\ref{sec:sims}, the synthetic profiles from these models generally fail to reproduce some of the chromospheric line shapes, particularly the single-peaked profiles in \cii, \mgii\ and \caii, so we carried out only a qualitative comparison with the observations (e.g., in terms of Doppler shifts).

Magnetic reconnection is generally accepted to play a significant role in small heating events, especially in the core of active regions. The specific properties of these events, and, for instance, whether they behave like scaled-down versions of larger flares, is highly debated. In this context, 
the study of accelerated particles in these events can shed light into the nature of the heating mechanisms and of particle acceleration processes as well. 
Recent works have revealed the presence of accelerated particles in very small heating events, although typically these are characterized by smaller low-energy cutoff and steeper slopes ($E_{\rm C} \sim 5$--15~keV, $\delta \gtrsim 7$) compared with larger flares (e.g., \citealt{Hannah2008,Testa2014,Wright2017,Testa2020,Glesener2020,Cooper2021}; see also review of \citealt{Testa_Reale_2022arXiv}, and references therein). These findings were based on diagnostics from direct detection of non-thermal particle emission at hard X-ray wavelengths (e.g., with RHESSI, FOXSI, NuSTAR; \citealt{Lin2002,Glesener2016,Harrison2013}), or spectral diagnostics from the indirect detection of the effect of non-thermal particles in the lower atmosphere where they can deposit most of their energy (e.g., with \iris, and with ground-based data such as the \sst\ data analyzed here). 
Recent work by \cite{Polito2023} presented rare simultaneous observations with \iris\ and NuSTAR, which provided independent diagnostics of non-thermal particles and found concordant results from both approaches. Also in this latter case the non-thermal particle distribution was characterized by relatively low energies ($E_{\rm C} \sim 8$~keV) and steep slopes, analogous to the properties of the heating event we studied here, as inferred from the comparison of \iris\ and \sst\ spectra with RADYN simulations.

The observations analyzed here put tight constraints on the models, but also highlight the shortcomings of the models. In fact, although they provide a good match for the observed TR \siiv\  profiles (and occasionally reproduce some chromospheric spectral profiles, such as the H$\alpha$ case discussed in Section~\ref{sec:sims}), they generally fail at producing chromospheric spectral line shapes similar to the observations as also discussed in previous work \citep[e.g.,][]{Polito2018,Testa2020,Cho2023}. We note that these discrepancies between models and observed profiles occur also in quiescent atmosphere (see e.g., \citealt{Hansteen2023}, and references therein).
A necessary next step is therefore to investigate in detail the cause of these discrepancies and improve the model to remediate them, and allow us to fully exploit these diagnostics.

\begin{acknowledgements}
We thank the referee for a careful review of the paper and several suggestions that have helped improve the paper.
We thank Vanessa Polito for the use of previously published RADYN simulations.
PT is funded for this work by contracts 8100002705 (\iris),
and NASA contract NNM07AB07C (\hinode/XRT) to the Smithsonian Astrophysical Observatory, and by NASA grant 80NSSC20K1272. 
BDP is supported by NASA contract NNG09FA40C (\iris).
This research has made use of NASA's Astrophysics Data System and of the SolarSoft package for IDL.
\sdo\ data were obtained courtesy of NASA/\sdo\ and the AIA and HMI science teams.
\iris\ is a NASA small explorer mission developed and operated by LMSAL with mission operations executed at NASA Ames Research Center and major contributions to downlink communications funded by ESA and the Norwegian Space Centre.
The Swedish 1-m Solar Telescope is operated on the island of La Palma by the Institute for Solar Physics of Stockholm University in the Spanish Observatorio del Roque de los Muchachos of the Instituto de Astrof{\'\i}sica de Canarias.
The Institute for Solar Physics is supported by a grant for research infrastructures of national importance from the Swedish Research Council (registration number 2021-00169).
This research is supported by the Research Council of Norway, project numbers 250810, 325491, and through its Centres of Excellence scheme, project number 262622.    
\end{acknowledgements} 


\end{document}